\documentclass[Afour,sageh,times]{sagej}
\usepackage[T1]{fontenc}
\usepackage{amsmath}
\usepackage{booktabs}
\usepackage{caption}
\usepackage{subcaption}
\usepackage{url}

\usepackage{graphicx}
\theoremstyle{thmstyleone}%
\theoremstyle{thmstyletwo}%

\theoremstyle{thmstylethree}%

\newcommand{\blue}{}
\newcommand\BibTeX{{\rmfamily B\kern-.05em \textsc{i\kern-.025em b}\kern-.08em
T\kern-.1667em\lower.7ex\hbox{E}\kern-.125emX}}

\def\volumeyear{2016}

\begin{document}
\runninghead{Karp et~al.}

\title{Experience and Analysis of Scalable High-Fidelity Computational Fluid Dynamics on Modular Supercomputing Architectures}

\author{Martin Karp\affilnum{1}, Estela Suarez\affilnum{2,3}, Jan H. Meinke\affilnum{2}, M\r{a}ns I. Andersson\affilnum{1}, Philipp Schlatter\affilnum{4,5}, Stefano Markidis\affilnum{1}, and Niclas Jansson\affilnum{6}}

\affiliation{\affilnum{1}Department of Computer Science, KTH Royal Institute of Technology, Stockholm, Sweden\\
\affilnum{2}J\"{u}lich Supercomputing Centre, Institute for Advanced Simulations, Forschungszentrum J\"{u}lich GmbH,  J\"{u}lich, Germany\\
\affilnum{3}Institute of Computer Science, University of Bonn, Bonn,  Germany\\
\affilnum{4} Institute of Fluid Mechanics (LSTM), Friedrich-Alexander Universit\"{a}t (FAU) Erlangen-Nürnberg,  Germany\\
\affilnum{5}SimEx/FLOW, Engineering Mechanics, KTH Royal Institute of Technology, Stockholm, Sweden \\
\affilnum{6} PDC Centre for High Performance Computing, KTH Royal Institute of Technology, Stockholm, Sweden}

\corrauth{Martin Karp, KTH Royal Institute of Technology, School of Electrical Engineering and Computer Science, Lindstedsv\"{a}gen 5, 100 44 Stockholm, Sweden}

\email{makarp@kth.se}
\begin{abstract}
 The never-ending computational demand from simulations of turbulence makes computational fluid dynamics (CFD) a prime application use case for current and future exascale systems. High-order finite element methods, such as the spectral element method, have been gaining traction as they offer high performance on both multicore CPUs and modern GPU-based accelerators. In this work, we assess how high-fidelity CFD using the spectral element method can exploit the modular supercomputing architecture at scale through domain partitioning, where the computational domain is split between {\blue a Booster module powered by GPUs and a Cluster module with conventional CPU nodes}. We investigate several different flow cases and computer systems based on the {\blue Modular Supercomputing Architecture} (MSA). We observe that for our simulations, the communication overhead and load balancing issues incurred by incorporating different computing architectures {\blue are} seldom worthwhile, especially when I/O is also considered, but when the simulation at hand requires more than the combined global memory on the GPUs, utilizing additional CPUs to increase the available memory can be fruitful. We support our results with a simple performance model to assess when running across modules might be beneficial. {\blue As MSA is becoming more widespread and efforts to increase system utilization are growing more important our results give insight into when and how a monolithic application can utilize and spread out to more than one module and obtain a faster time to solution.}
\end{abstract}

\keywords{Computational Fluid Dynamics, Modular Supercomputing Architecture, HPC.}
\maketitle              %
\section{Introduction}

Computational fluid dynamics (CFD) impacts many fields ranging from medicine to aeronautics and is one of the largest application domains in modern HPC systems~\citep{slotnick2014cfd}. Designing efficient CFD software tailored to the most powerful supercomputers is an active area of research and developing methods and algorithms that map to upcoming heterogeneous hardware is growing ever more important~(\citealt{abdelfattah2021gpu}). 

The modular supercomputing architecture (MSA) is uniquely positioned as one of the main enabling technologies for the European exascale computer ecosystem. It combines different modules tailored for specific sets of algorithms and applications connected with a high-performance interconnect. This type of supercomputing cluster provides a dynamic and flexible system for a wide range of applications and use cases~(\citealt{suarez2019modular,kreuzer2021porting}). It has already been deployed in both the JURECA and JUWELS supercomputers at J\"{u}lich Supercomputing Centre (JSC) and is posed to be the computing architecture for a future exascale computer system at JSC~(\citealt{krause2018jureca,krause2019juwels}). However, applications need to be adapted to take advantage of more than one module at a time.

Through dedicated efforts, MSA has already been accommodated in several applications such as multiphysics or multiscale applications that can efficiently run large well-defined code sections on different computing modules~(\citealt{kreuzer2021porting,riedel2021practice,markov2019large}). By splitting the code execution and running large parallel regions on the Booster modules dedicated to energy-efficient high-throughput processing units such as GPUs and running portions with low scalability on the cluster module focused on providing low latency and high frequency, a large improvement in the performance of the solver has been observed~(\citealt{Kreuzer:851724}). However, some application domains are dominated by large homogeneous "monolithic" solvers where each process executes the same operations and only the computational domain is partitioned.

The benefits of MSA for these types of applications, which occur in various domains revolving around solving one large partial differential equation such as solid mechanics or fluid dynamics, are less clear-cut. On a large scale, when the problem will not fit on any one module, it comes down to distributing the work between different modules appropriately. For smaller problems, it instead becomes an issue of choosing the most suitable module to execute the computation on. {\blue As flexible job scheduling is becoming more important to increase system utilization, understanding the performance implications of using multiple modules for these types of applications is also becoming more relevant~\citep{arima2022convergence}. One aspect of this work is to assess when utilizing several modules can reduce the time to solution for scalable monolithic solvers.}

In this work, we evaluate how large-scale {\blue high-fidelity} computational fluid dynamics simulations {\blue based on solving the Navier-Stokes equations can utilize different MSA modules at the same time and how workloads of different sizes are best run on a heterogeneous MSA system. High-fidelity CFD makes up a large share of the computational load on many supercomputers, and due to the  demand for more grid points and higher resolution, there is a never-ending need for computational resources. This approach differs from lower-fidelity models such as the Reynolds-averaged Navier-Stokes or other approaches more suited for complex geometries such as Lattice-Boltzman, where the Boltzmann equations are solved instead.} We use a CFD solver that performs well on both CPUs and GPUs combined with a simple performance model to analyze and understand how we distribute a workload and execute computations on two different MSA systems, the JUWELS cluster and Booster modules as well as the DEEP cluster and booster modules. We claim the following contributions:
\begin{itemize}
    \item We empirically compare different flow configurations across different GPU/CPU configurations, utilizing not only GPUs and CPUs but also mixing the two architectures on MSA. We also evaluate the impact of I/O on the load balance.
    \item We employ a simple performance model to reason about our results and evaluate the performance potential by running on multiple architectures.
    \item When the simulation cannot fit on the GPU module only, by using both GPU and CPU modules, we observe up to $2.7\times$ improved performance than only using the CPU module on the DEEP prototype system. We also compare the performance between the JUWELS Booster and LUMI-G module.
\end{itemize}

\section{Related Work}
{\blue This work relates both to various applications utilizing multiple modules on MSA, as well as CFD in general on heterogeneous computer architectures. While most efforts for CFD have been spent on optimizing the code for systems where the nodes internally are heterogeneous, our work explores how a solver optimized for different types of nodes can run using multiple compute modules with different node architectures by partitioning the computational domain between the different modules.

\subsection{CFD on Heterogenous Architectures}}
In the era of heterogeneous platforms, high-order methods for CFD have been gaining increasing amounts of interest {\blue for high-fidelity CFD due to their accuracy, structure, and relatively high number of floating point operations per grid point which }enable them to efficiently utilize GPUs in addition to multicore CPUs~(\citealt{abdelfattah2021gpu}). In the development of these methods, the focus has been on offloading the computation to the accelerator and limiting the data exchange from the host to the device as far as possible.
 
In this paper, to assess the performance of mixing different architectures, we consider a spectral element solver, {\blue Neko, running on nodes composed of CPUs as well as nodes powered primarily by GPUs with a host CPU. } Neko uses modern Fortran together with hand-written CUDA/HIP kernels behind a device abstraction layer to provide tuned implementations for all the different architectures~(\citealt{jansson2021neko,karp2022large,jansson2023exploring}). {\blue While there are many other methods to carry out fluid simulations, we focus on the Neko application, which integrates the Navier-Stokes equations in time and is able to efficiently scale using domain decomposition. CFD can take many forms on heterogeneous computer architectures, ranging from compressible solvers~\citep{witherden2014pyfr} to Lattice-Boltzman methods~\citep{calore2019optimization} and many others \citep{niemeyer2014recent}. However, not all solvers scale to the same extent as Neko and can utilize different computer architectures at a high parallel efficiency. For our work on high-fidelity CFD running on large-scale heterogeneous architectures, the spectral element method (SEM) is a good representative, and two SEM codes were because of this recently considered for the Gordon-Ball prize~\citep{jansson2023exploring,merzari2023exascale}.

There are many approaches targeting CFD, utilizing both CPUsand GPUs, as there are also different ways of utilizing mixed CPU-GPU nodes. Within a node, some approaches try to either offload certain tasks to the host CPU~(\citealt{borrell2020heterogeneous,calore2019optimization})}, or partition the computational domain between computing devices depending on their respective performance~(\citealt{zhong2014data,alonazi2015design,liu2016hybrid}). In our work, we are concerned with the second approach, {\blue but with the difference that we split the domain between two different computer modules}.  The works by \cite{zhong2014data,alonazi2015design} indicate that partitioning the domain between different computing devices can lead to improved performance, but this is in practice not done in many large-scale CFD solvers~(\citealt{abdelfattah2021gpu,kolev2021efficient}) {\blue because data movement between the CPU and GPU quickly becomes the limiting factor}. Our work aims to assess why and when a CFD application should consider using a mixture of different computing modules, assuming optimal load balancing.  
We are the first, to our knowledge, to study the performance of a CFD code for large scale production runs on a mix of compute modules with hundreds of GPUs or thousands of cores. {\blue The motivation of this work is first to enable running large-scale monolithic solvers such as Neko across compute modules when the HPC cluster is underutilized, and second to determine from the application point of view when mixing modules is compelling for actual production cases.

\subsection{Applications on MSA}
Different applications have been tested on the MSA. In particular, large performance improvements have been made possible for applications employing coarse-grained parallelism, in which different parts of the code benefit from different computer architectures and only limited communication between the compute modules is necessary. Notable examples are the implicit particle in cell method in xPIC by \cite{Kreuzer:851724} and machine learning \cite{riedel2021practice}. Further approaches across a wide range of applications reported in \cite{kreuzer2021porting}. However, as mentioned, the primary focus has been on dedicating specific computational resources to code parts with very different computational characteristics. 

Our code, on the other hand, simulates an incompressible flow that lacks coarse-grained isolated tasks; instead, we partition the domain between different computing devices. Going forward we see an opportunity for workflows where several coarse-grained tasks are executed in parallel, in addition to the actual simulation. One such approach, where in-situ data analysis is executed in parallel to the Neko simulation is suggested by~\cite{ju2023situ}. While we focus on domain-partioning in this paper, considering such approaches, and for example running the in-situ data analysis on a different module than the simulation is a natural extension to this work.}

\section{Computational Fluid Dynamics in HPC}
Fluid dynamics has been one of the focus areas of high-performance computing since its conception. Due to the vast array of application areas such as medicine, aerodynamics, and weather and climate models, detailed simulations of flows are of large scientific interest. High-fidelity simulations of the turbulent Navier-Stokes equations require tremendous computing power and a very fine resolution making them prime candidates for taking advantage of large, modern HPC systems. In this work, we focus on the integration in time of the non-dimensional incompressible Navier-Stokes, described by
\begin{equation}\label{eq:incom}
\begin{split}
    \nabla  \cdot  \mathbf{v} &= 0, \\ 
    \frac{\partial \mathbf{v}}{\partial t} + (\mathbf{v}\cdot \nabla) \mathbf{v} &= - \nabla p + \frac{1}{Re}\nabla^2 \mathbf{v} + \mathbf{F},
\end{split}
\end{equation}
where $\mathbf{v}$ is the instantaneous velocity field, $p$ the pressure, $Re$ is the non-dimensional Reynolds number and $\mathbf{F}$ an external forcing. The Reynolds number is defined as $Re=\frac{LU}{\nu}$ where $U$ is a characteristic velocity, $L$ is a suitable length scale, and $\nu$ is the kinematic viscosity. The Reynolds number is important in this context as a single direct numerical simulation of these equations, where all the scales of the flow are resolved, requires a grid that scales as $\mathcal{O}(Re^{9/4})$ for isotropic, homogeneous turbulence. This means that direct numerical simulation at even moderately high Reynolds numbers is extremely expensive. {\blue While there are many other approaches to CFD, our focus is on the integration of the Navier-Stokes equations in time with low numerical dispersion and high scalability. In our context, methods such as SEM are the prime candidates~\citep{deville2002high}.}
\subsection{Neko}
To assess how high-fidelity CFD simulations can be efficiently performed on varying computer hardware we will be utilizing Neko~(\citealt{jansson2021neko}), a Navier-Stokes solver based on the spectral element method. It has its roots in the long-running solver, Nek5000~(\citealt{nek5000}), which has scaled to over a million MPI ranks and was awarded the Gordon Bell price in 1999~(\citealt{tufo1999terascale}). Neko provides the same excellent scaling capabilities as Nek5000 on modern multicore systems and adds support for more recent computer architectures such as GPUs~(\cite{karp2022large}). This makes it a suitable candidate to assess how we can leverage a wide range of different computer architectures for large CFD simulations.

While several other methods are used for CFD, not all can utilize GPUs efficiently or scale to a large number of MPI ranks. Oftentimes a low operational intensity, the number of floating operations executed per byte, and the prevalence of complex global communication patterns make it difficult to utilize massively parallel architectures such as GPUs. Our choice of discretization and solver relates to this: the spectral element method has shown major promise in enabling CFD simulation at the exascale due to its high-order and local structure, enabling efficient utilization of both CPUs and GPUs~(\citealt{fischer2020scalability,abdelfattah2021gpu}).

Due to the globally unstructured but locally structured nature of the spectral element method, only unit-depth communication is necessary in a so-called gather-scatter phase~(\citealt{deville2002high}). All other operations can be performed in an element-by-element or matrix-free fashion, which yields a high level of parallelism and utilizes both multicore CPUs and GPUs efficiently. At the heart of the method, similar to many other CFD solvers, preconditioned Krylov subspace methods are used to solve linear systems on the form $Ax=b$ for each time step.
The exact splitting of the velocity and pressure follows a similar splitting as outlined by \cite{karniadakis1991high} and described for Neko in \cite{karp2022large}. For the resulting linear systems, we use restarted GMRES for the pressure solves with a hybrid-Schwarz multigrid preconditioner, while for the velocity we use CG together with a block-Jacobi preconditioner. While there are other pipelined Krylov methods and implementations available in Neko~(\citealt{karp2022reducing}), for this study we evaluate the original and most common configuration.

In the spectral element method, the computational domain is split into $E$ non-overlapping hexahedral elements. These parts of the domain are then distributed among the MPI ranks and it is through this domain partitioning that the spectral element method leverages the parallelism of modern computing architectures. The flow field is represented on the reference element with high-order polynomial basis functions of order $N$, collocated on the Gauss-Lobatto-Legendre points and is desrbied extensively in \cite{deville2002high}. The computational load is identical for each element. The only asymmetry that is introduced is through the gather-scatter operation, which depends on the geometric distribution of the elements across the MPI ranks.
 
\subsection{Flow Cases under Consideration}\label{sec:cases}
With our focus on high-fidelity simulations of turbulent flow, we consider three different simulation cases of varying sizes. We summarize the details of each flow case in Table~\ref{tab:cases}. We use a polynomial order of $N=7$ as most simulation cases use a polynomial order between 5 and 11.

\begin{table*}[]
    \centering
    \begin{tabular}{llrr}\toprule
        Case & $N$ & $E$  & $n=EN^3$\\\midrule
         Turbulent pipe $Re_b=5300$& 7 & 36~480  &12~512~640\\
         Taylor-Green vortex $Re=1600$& 7 &  262~144  &89~915~392\\
         Rayleigh-B\'{e}nard convection, $Ra=10^{11}$ ~&7 & 2~097~152& ~719~323~136 \\\bottomrule
    \end{tabular}
    
    \caption{Flow cases under consideration. Polynomial order $N$, number of elements $E$, and total number of unique grid points, $n$.}
    \label{tab:cases}
\end{table*}

\subsubsection{Turbulent Pipe.}
Turbulent flow in a pipe is a canonical flow case, which occurs in biological applications such as blood flow, and industrial applications such as gas and oil pipelines. One case that has been studied extensively, is the flow in a turbulent pipe at bulk Reynolds number $Re_b=5300$ based on the cylinder diameter and bulk flow velocity $U_b$. We consider this case as a smaller simulation case, only requiring a few nodes to efficiently compute. The exact details of the flow case are described by~\cite{el2013direct}.

\subsubsection{Taylor-Green Vortex.}
The Taylor-Green vortex (TGV) has been studied extensively in order to assess the accuracy and convergence of CFD solvers. In the TGV case, the Reynolds number is uniquely defined by the viscosity and in particular, TGV at $Re=1600$ has been used previously~(\citealt{VANREES20112794}). We use this case to assess the scaling behavior of a medium-sized workload requiring a moderate number of nodes to execute efficiently.

\begin{figure}
    \centering
    \includegraphics[width=0.33\columnwidth]{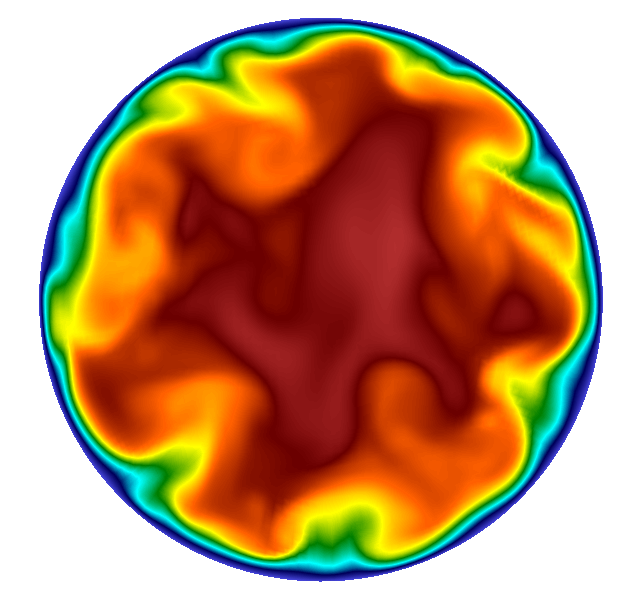}
    \includegraphics[width=0.31\columnwidth]{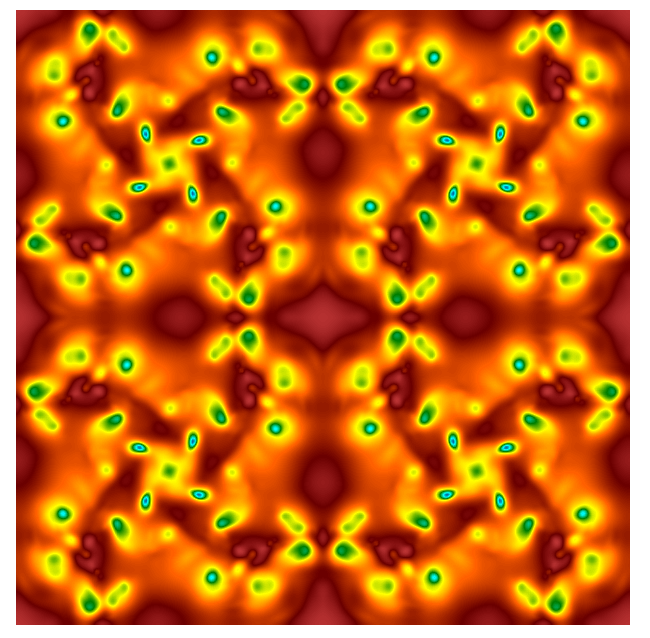}
    \includegraphics[width=0.3\columnwidth]{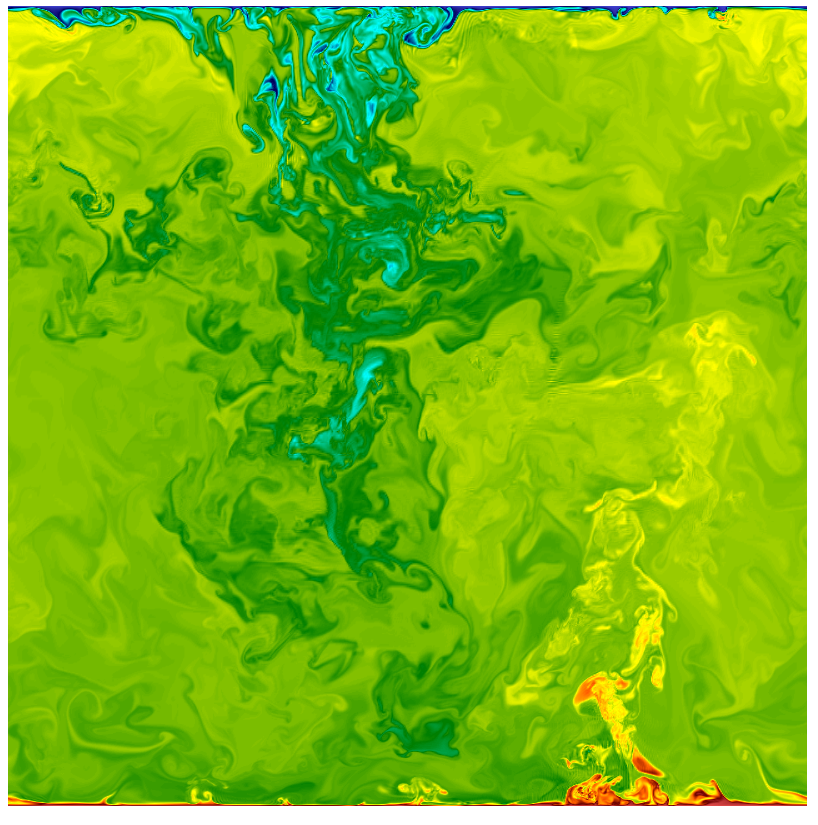}
    \caption{Visualizations of the three different cases, with red being high and blue being a lower value. To the left is the velocity magnitude in a cross-section of the pipe, in the middle is the pressure field in TGV and to the right, we show the temperature field in turbulent Rayleigh-B\'{e}nard convection.}
    \label{fig:rbc}
\end{figure}
\subsubsection{Rayleigh-B\'{e}nard Convection.} For our largest case we consider Rayleigh-B\'{e}nard convection (RBC), which simulates the same physical behavior that occurs in the sun and many industrial applications \citep{iyer2020classical} where the increased buoyancy of a hotter fluid drives convective turbulence as shown in Figure \ref{fig:rbc}. In this work, we consider a cubic domain with an aspect ratio of $1$, periodic sides, and walls on the top and bottom where the bottom wall has a temperature of 1. We perform this simulation at a Rayleigh number of $10^{11}$ and a Prandtl number of 1. Our simulation follows a similar setup to the cubic case in \cite{kooij2018comparison}, but at a higher Rayleigh number. As this case is rather large we want to consider how to utilize several modules when one module might be too small to fit the entire problem. 

\section{Performance Analysis}
In this section, we perform a performance analysis where we relate the performance and memory capacity of different computing devices to reason around when and how it might be beneficial to split a homogeneous problem, where each device performs the same task on different parts of the problem, across different computing devices and supercomputer modules. We first develop a simple model to reason around the performance of mixing different computing devices and then go on to identify different domains of operation for a homogeneous workload, in what domains our performance model will work well, and {\blue what performance improvements one can expect in the best case by using different compute modules.}

\subsection{Performance Model for Mixing Different Computing Devices}
We develop a simple performance model for computations revolving around solving one large system by splitting a homogeneous computational cost (such as the computational domain) between different computing units (such as GPUs and CPUs). The aim of this model is to provide an optimistic indication of when using several computing modules might be beneficial, not to predict the exact run time of an application. The model is similar to what was originally proposed by Amdahl and similar to what has been used previously to discuss the performance and scalability of PDE solvers~(\citealt{fischer2015scaling}). 

\begin{figure}
\centering
 \begin{subfigure}[b]{0.49\textwidth}
\centering
    \includegraphics[width=\textwidth]{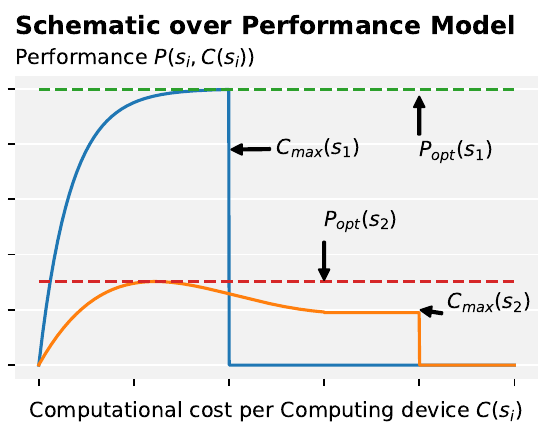}
    \caption{}\label{fig:model_perf}
 \end{subfigure} 
 \hfill
 \begin{subfigure}[b]{0.49\textwidth}
    \centering
    \includegraphics[width=\textwidth]{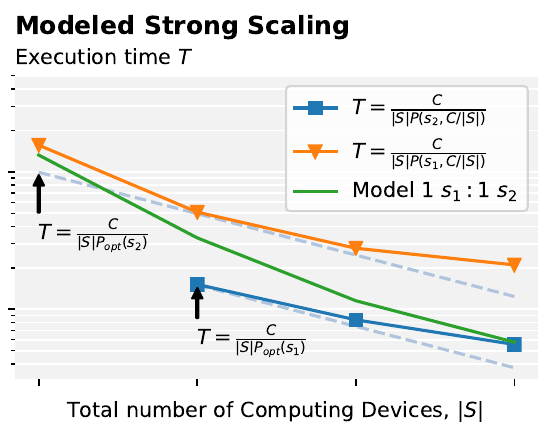}
    \caption{}\label{fig:model_strong}
 \end{subfigure}
 \caption{Illustration of the performance model for two different computing devices $s_1,s_2$ with different performance characteristics. We denote the modeled time as described in \eqref{eq:model} as Model $1~s_1:1~s_2$ and we model the best achievable performance based on $P_{opt}(s_1),P_{opt}(s_2)$ with a mix of 1:1, $s_1,s_2$ devices. The strong scaling performance for $|S|$ computing devices with a performance based on Figure \ref{fig:model_perf} is shown as in Figure \ref{fig:model_strong}.}\label{fig:perf_model}
\end{figure}

We denote the execution time of a simulation with $T$ and divide it into two non-overlapping sections: 
\begin{equation}
T=T_a+T_c,
\end{equation}
where $T_a$ is the local time dedicated to arithmetic operations and loads and stores to and from global memory (DRAM or high bandwidth memory (HBM)), while the communication time $T_c$ is the latency portion of the run time that is used for communication between different MPI ranks and inherent latency of the computing devices. We also introduce the computational cost or work $C$ for a given workload which is then divided among all computing devices $C = \sum_{s_i \in S} C_i$. Each computing device $s_i \in S$, where $S$ is the set of computing devices, then has a performance $P(s_i,C_i)$ given that computing device $s_i$ is computing a cost of $C_i$. The units for $C,P$, depend on the problem, but in our case the cost is related to the computation of one time step, meaning that the cost is given in time steps and the performance in time steps per second. For a given processing device $s_i$ computing a cost $C_i$ we have that 
\begin{equation}
    T_a(s_i,C_i)=\frac{C_i}{P(s_i,C_i)}.
\end{equation} 
What we would like to obtain is the minimal run time overall computing devices, and hence solve the minimization problem
\begin{equation}
\begin{split}
\underset{C_i}{\text{minimize }} &T = \max_{s_i \in S} (T_a(s_i,C_i) + T_c(s_i)) \\
\text{such that. }&T_a(s_i,C_i)=\frac{C_i}{P(s_i,C_i)}\\
&C = \sum_{s_i \in S} C_i\\
&C_i \le C_{max}(s_i), \quad s_i \in S
\end{split}  
\end{equation}
where we introduce the capacity of computing device $s_i$ as $C_{max}(s_i)$, which is the largest cost a given computing device can compute, often limited by e.g. DRAM or HBM memory capacity. For our model, we focus on finding a lower bound on the run time and comparing the results of our performance measurements to this optimistic lower bound. To do this, we start by observing that $T \ge T_a$ and as such we can trivially lower bound the performance and run time for the computing device as
\begin{equation}
    T \ge \frac{C_i}{P(s_i,C_i)} \ge \frac{C_i}{P_{opt}(s_i)} 
\end{equation}
where we introduce $P_{opt}(s_i)$, which corresponds to the highest performance achievable for processing device $s_i$. With this information, we can provide a lower bound on the lowest possible run time $T_{min}$ as
\begin{equation}
T_{min} \ge \max_{s_i\in S} \frac{C_i}{P_{opt}(s_i)}
\end{equation}
subject to the constraint that $C=\sum_{s_i\in S} C_i$. For the unconstrained case, when all computing devices have enough memory to fit their part of the cost $C$, this reduces to
\begin{equation}\label{eq:model}
    T_{min} \ge \frac{C}{\sum_{s_i\in S} P_{opt}(s_i)}
\end{equation}
and the relation  $C_i/P_{opt}(s_i) = C_j/P_{opt}(s_j), \forall s_i, s_j \in S$ holds. In the other case, we have that there exists some computing devices s.t. $$C_{max}(s_i)/P_{opt}(s_i) < C_j/P_{opt}(s_j), s_i, s_j \in S$$ and the optimization problem does not necessarily have a simple solution. As we consider only two different computing devices in this work (one kind of GPUs and CPUs used at the same time), solving this problem is not an issue, but if the performance $P(s_i)$ would vary significantly among the computing devices $s_i \in S$, the number of constraints would increase considerably. To summarize, our modeled lowest possible run time of our mixed GPU/CPU runs is computed as the following:
\begin{equation}\label{eq:model}
\begin{split}
    \text{If }~&\frac{C_i}{P_{opt}(s_i)} = \frac{C_j}{P_{opt}(s_j)},\quad \forall s_i, s_j \in S,\\
              &C = \sum_{s_i \in S} C_i, \\
              &C_i \le C_{max}(s_i), \quad s_i \in S, \\
              \text{then: } ~&T_{min} = \frac{C}{\sum_{s_i\in S} P_{opt}(s_i)}\\
    \text{Else: } ~&\\
    \underset{C_i}{\text{minimize }} &T_{min} = \max_{s_i\in S} \frac{C_i}{P_{opt}(s_i)}\\
              \text{such that. } &C = \sum_{s_i \in S} C_i\\
               &C_i \le C_{max}(s_i), \quad s_i \in S \\
\end{split}
\end{equation}
The best case is that the performance of two computing devices is additive if they can fit the entire problem. Another takeaway from this model is that we can achieve significant superlinear speedup when a single module of computing devices cannot hold the entire computational cost and we are limited by the capacity of the devices. Increasing the capacity then effectively yields a superlinear speedup until the modules can hold enough of the computational work. We illustrate the meaning of our notation in Figure \ref{fig:perf_model}, for a simple case with two different computing devices, $s_1,s_2$. Given the single node performance shown in (a), the modeled performance as we scale is shown in (b).

For Neko, we let the cost $C$ be a linear function of the number of elements, $E_i$, on a computing device and model the performance according to equation \ref{eq:model}. As such, finding $T_{min}$ can be done through a parameter search where we load balance the elements between the different computing devices. The best performance $P_{opt}(s_i)$ for the GPUs and CPUs is approximated as the best-measured performance for a given flow case, using only CPUs/GPUs. We visualize the modeled time with a solid line in our experimental results along with our mixed GPU/CPU runs, similar to the modeled strong scaling in Figure \ref{fig:perf_model}. A similar approach can be applied to any other solver solving one large problem through domain partitioning.

\subsection{Operation Domains}
An aspect of the modeled time that we propose is that we do not consider the communication time $T_c$, but we assume that the whole problem scales perfectly. This is most often not the case, but it depends strongly on the problem size, and thus the cost per computing device $C_i$ which relates to the relation between $T_a$ and $T_c$. As such, we introduce three different domains of operation for an application with different performance characteristics, where the computation is either dominated by $T_a$ or $T_c$, and discuss where running on a mix of computing devices might be beneficial.

\begin{equation}
    \begin{alignedat}{2}
        T_a &\leq T_c, &\quad\text{\blue Communication domain}\\
        T_a & > T_c,&\quad\text{Scaling domain}\\
        T_a & \gg T_c, \quad C \approx C_{max}, &\quad\text{Extreme scaling domain}.\\
    \end{alignedat}
\end{equation}
In the {\blue communication domain}, it does not make sense to add computational resources, as $T_c$ in general increases with the number of processing devices, and we are already limited by communication (latency). In this domain, CPUs may have an edge due to their low latency concerning memory and communication and high clock speeds.  This is the case for many applications, which do not have the opportunity to scale on GPUs or to a large number of nodes and this is the domain the Cluster module caters to. 

In the scaling domain, the total amount of work is still the dominating factor for the application performance, hence adding more compute units would be beneficial. However, in this domain, it is still not evident that we will easily be able to balance the different computing units in such a way that we get a reduction in run time. However, as we are primarily limited by computational power, throughput-oriented devices such as GPUs tend to be the most performant and power-efficient option, which is the idea behind the Booster module~(\citealt{kreuzer2021porting}).

In the extreme scale regime, we are considering examples in which the computational cost $C$ is close to the capacity $C_{max}$ of the available resources and might not fit into any single compute module. In this situation, the ability to use several modules to fit a large case becomes crucial, which justifies the potential loss in workload balance. The total performance, assuming $T_c$ is small, will overall be additive and follow our performance model, and the major appeal is that cases that are impossible to run otherwise will now be possible. Overall, these cases would then not treat the Cluster and Booster modules as two different modules, but rather as two pillars to compute these extremely large systems. This domain most closely correlates with our proposed performance model, while the model would provide optimistic performance bounds in the first two domains.

Neko, similarly to many flow solvers is primarily memory bound for the computational cost $C$, while the communication overhead, $T_c$, can be primarily attributed to the gather-scatter kernel. This is consistent with previous works where the gather-scatter kernel has shown to be the main {\blue performance bottleneck of SEM as one approaches the strong scaling limit}, and a heavily optimized version is integral for high performance~(\citealt{ivanov2015evaluation}). The gather-scatter kernel is called repeatedly for each operator evaluation and has a strong dependence on the distribution of the work among the available ranks as it performs the unstructured communication among MPI ranks and elements.

\section{Experimental Setup}
In this work, our primary experimental platforms are based on the modular supercomputing architecture (MSA)~(\citealt{suarez2019modular}). MSA groups different kinds of compute nodes into sub-clusters (modules) that are internally rather homogeneous. The node architecture of each module targets the needs of a specific kind of application. Depending on the required network topology new modules can be added and extended easily. 

An example is the JUWELS supercomputer---one of the largest systems in Europe---at the J\"{u}lich Supercomputing Center. It currently accommodates two different computing modules (Cluster and Booster) that share a single high-performance interconnect. With this design, it is possible to dynamically map applications with vastly different performance characteristics to the modules and accommodate a wide range of use cases. The JUWELS Cluster is a CPU-based HPC system, good for applications (or parts of them) that are not ready to run on GPUs and/or require high single-thread performance. The Booster module utilizes GPUs and is used by the most scalable applications with high-performance demands.

The DEEP system, a prototype for the modular supercomputing architecture provides in addition to a cluster and a booster module, a module dedicated to data analytics. This module is equipped with large, fast, storage as well as  GPUs and FPGAs for extensive data processing. By sharing the same interconnect it is possible to assign different tasks to the modules that are executed in-situ while the simulation is running.

Aside from the two systems just described, we evaluate the LUMI supercomputer at CSC in Finland. While LUMI shares a similar modular architecture to the systems at JSC, with different modules for CPU and GPUs, the vast amount of the resources is dedicated to the GPU/Booster module LUMI-G, which we will consider. We focus on the three production use cases described in subsection~\ref{sec:cases} to capture actual production usage and do not evaluate any proxy app or similar, but the whole application. For all measurements we use a shaded area to indicate the 95\% confidence interval for the time of any time step of the simulation, assuming that the time per time step follows a normal distribution around the sample mean. We use the last 100 time steps of each simulation to collect our performance measurements.

We provide an overview of the different computational setups and the two modules of each that we use in Table~\ref{tab:exp_setup}. A major difference from LUMI-G as compared to the Booster module of JUWELS is that the network interface cards (NIC) are mounted directly on the GPUs, essentially offloading also the communication in addition to the computation to the GPU. {\blue On JUWELS, in comparison, the Mellanox HDR200 is connected to the GPUs through a PCIe switch that is shared with the host CPUs. The topology of the networks in the computers also differs: LUMI-G is arranged in a more conventional Dragonfly topology \citep{kim2008technology}, while JUWELS uses a Dragonfly+ network topology as proposed by \cite{shpiner2017dragonfly+}.}
\begin{table*}[]
    \centering
    \begin{tabular}{lll}\toprule
         \bf JUWELS & \bf Cluster & \bf Booster  \\ \midrule
         Compute nodes &  2271 & 936\\
         CPU & 2 $\times$ 24 core Intel Xeon 8168 &2 $\times$ 24 core AMD EPYC 7402\\
         CPU Memory & 96~GB DDR4-2666 RAM &512 GB DDR4-3200 RAM\\
         GPU & - & 4x Nvidia A100\\
         GPU Memory & - &40 GB HBM \\
         Interconnect & Mellanox InfiniBand EDR100 & 4 $\times$ Mellanox HDR200 InfiniBand \\
         Compiler & GCC 11.3.0 &GCC 11.3.0\\
         CUDA/ROCM &- & CUDA 11.7 \\
         MPI & OpenMPI 4.1.4 & OpenMPI 4.1.4\\ \toprule
        \bf DEEP & \bf Cluster & \bf Booster\\\midrule
         Compute nodes & 50 & 75 \\
         CPU & 2 $\times$ 12 core Intel Xeon 6146 & 2 $\times$ 8 core Intel Xeon 4215\\
         CPU Memory & 192~GB DDR4 & 48~GB DDR4\\
         GPU & -  & Nvidia V100 \\
         GPU Memory & - & 32~GB HBM\\ 
         Interconnect & Mellanox InfiniBand EDR100 & Mellanox InfiniBand EDR100 \\ 
          Compiler & Intel 2021.4.0 & Intel 2021.4.0 \\
          CUDA/ROCM &- & CUDA 11.7 \\
         MPI &ParaStationMPI 5.5.0 &  ParaStationMPI 5.5.0\\ \toprule
          \bf LUMI & &\bf LUMI-G  \\\midrule
          Compute nodes & &2560\\
         CPU & &1 $\times$ 64 core AMD EPYC 7A53\\
         CPU Memory & & 512 GB DDR4 \\
         GPU & &4 $\times$ AMD Instinct MI250X\\
         GPU Memory & &128GB HBM2e\\
         Interconnect &  &4 $\times$ 200GB/s Slingshot-11 \\
         Compiler &&CCE 14.0.2 \\
         CUDA/ROCM && ROCM 5.0.2 \\
         MPI & &cray-mpich 8.1.18\\
         \bottomrule
    \end{tabular}
    \caption{Software details and hardware details per node of the different computer modules and setups.}
    \label{tab:exp_setup}
\end{table*}
All runs in Neko are executed with one MPI rank per CPU core for the CPU nodes, and one MPI rank per logical GPU for the GPU nodes. For our experiments mixing GPUs and CPUs, we use Neko extended with support to distribute the number of elements unevenly between different MPI ranks. For the distribution of the elements we then first partitioned the mesh with ParMETIS~\cite{karypis2003parmetis} and after this, we performed a parameter search to find the best weight (how many elements each core/GPU should compute) between the GPU and CPU devices for each case. 

For the execution of the inter-module cases, we utilized the heterogeneous job scheduling available on JUWELS and DEEP. As we are comparing a wide range of computational platforms we introduce the notion of a computing device for a computational platform. We define the computing devices for each platform as one CPU node on DEEP/JUWELS or one logical GPU, meaning one Graphics Compute Die (GCD) of the MI250X or one V100/A100 GPU. We provide an overview in Table \ref{tab:compute_devices}. In our mixed runs, we use a mixture of one CPU computing device and one GPU computing device to illustrate the performance behavior when mixing computer modules.

We utilize LLView on the DEEP system to collect statistics for MSA runs with and without significant amounts of I/O. This is to identify how the workload and load balance changes if the  I/O load increases compared to the computational workload.

\begin{table}[]
    \centering
    \begin{tabular}{ll}\toprule
        Compute cluster & Computing Device\\\midrule
        DEEP-Booster & 1 V100 GPU\\
        DEEP-Cluster & 1 node with 2x12 Intel CPU cores\\
        JUWELS-Booster & 1 A100 GPU\\
        JUWELS-Cluster & 1 node with 2x24 Intel CPU cores\\
        LUMI-G & 1 MI250X GCD\\\bottomrule
    \end{tabular}
    \caption{List of computing devices used in the experiments. For the MSA runs we utilize a  1:1 Mix where one computing device from the Booster and Cluster module is used simultaneously.}
    \label{tab:compute_devices}
\end{table}

\section{Results}
In this section, we detail the performance measurements for the different simulation cases across the experimental platforms and discuss how the results relate to our previous performance analysis.
We show the standard deviation with a shaded area in all plots.

\subsection{Performance Measurements}
\begin{figure*}
\centering
 \begin{subfigure}[b]{0.49\textwidth}
\centering
    \includegraphics[width=\textwidth]{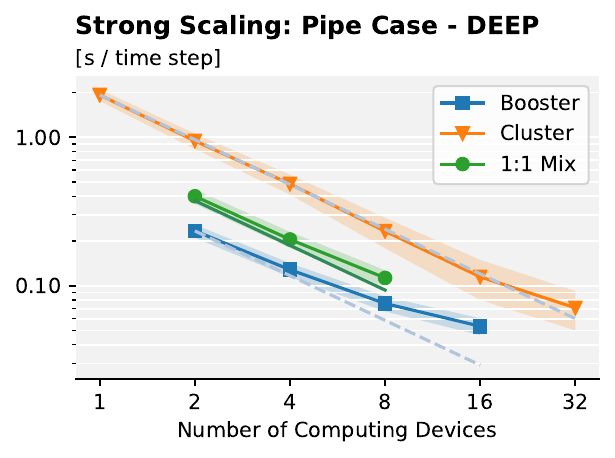}
    \caption{Performance on the DEEP supercomputer.}
    \label{fig:my_label}
 \end{subfigure} 
 \hfill
 \begin{subfigure}[b]{0.49\textwidth}
    \centering
    \includegraphics[width=\textwidth]{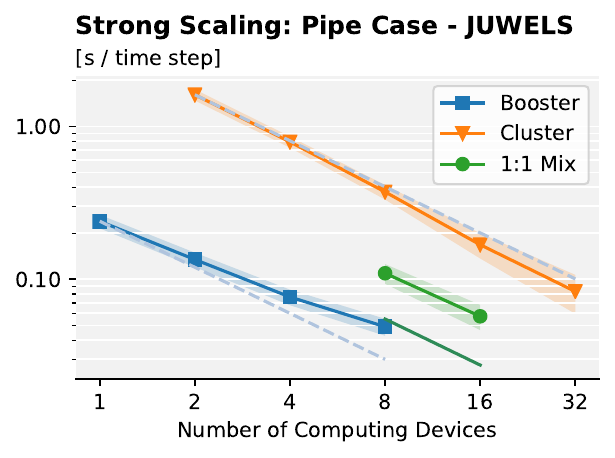}
    \caption{Performance on the JUWELS supercomputer.}
    \label{fig:RBC}
 \end{subfigure}
 \begin{subfigure}[b]{0.49\textwidth}
\centering
    \includegraphics[width=\textwidth]{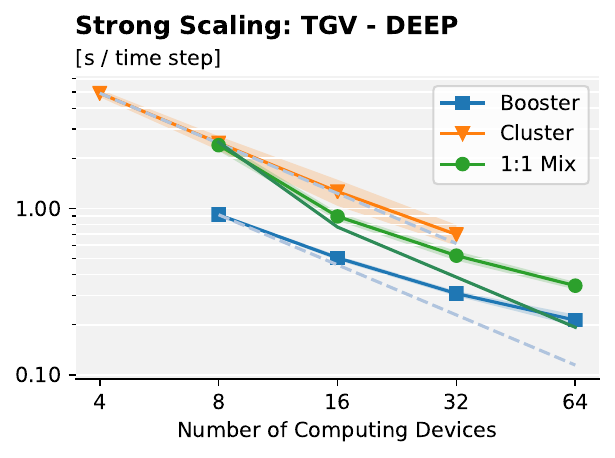}
    \caption{}
    \label{fig:my_label}
 \end{subfigure} 
 \hfill
 \begin{subfigure}[b]{0.49\textwidth}
    \centering
    \includegraphics[width=\textwidth]{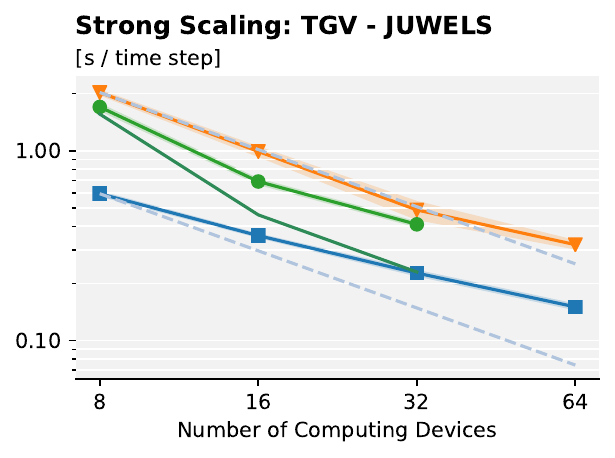}
    \caption{}
    \label{fig:my_label}
    \end{subfigure}
     \begin{subfigure}[b]{0.49\textwidth}
\centering
    \includegraphics[width=\textwidth]{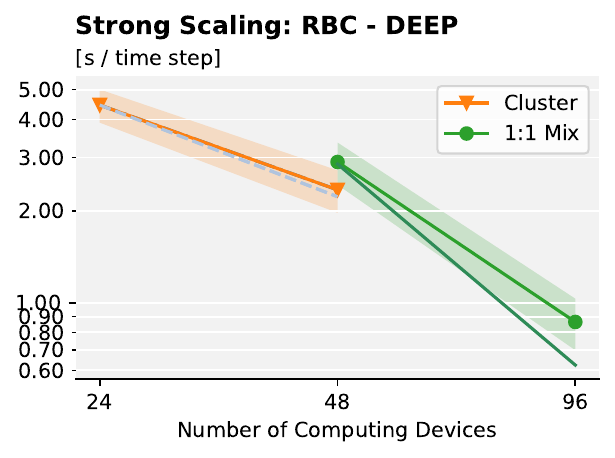}
    \caption{}
    \label{fig:my_label}
 \end{subfigure} 
 \hfill
 \begin{subfigure}[b]{0.49\textwidth}
    \centering
    \includegraphics[width=\textwidth]{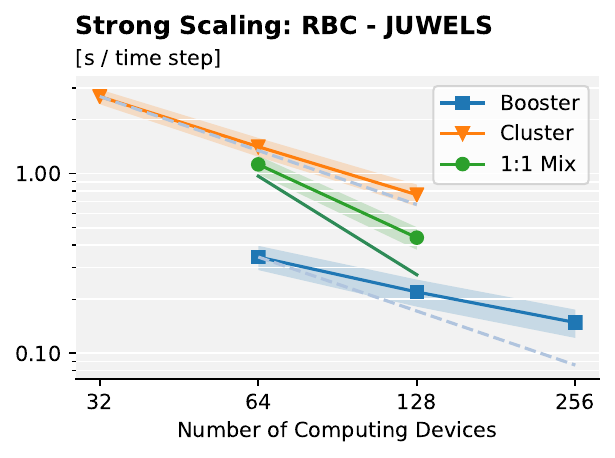}
    \caption{}
    \label{fig:my_label}
 \end{subfigure}
 \caption{Performance comparison between DEEP and JUWELS for our three different test cases. We show perfect linear scaling for the Booster and Cluster runs with a dotted line while we show the modeled performance with a green solid line without markers for the MSA runs. The modeled time is based on the highest performance $P_{opt}$ for the given case measured on the Booster and Cluster modules.}\label{fig:all_runs}
 \end{figure*}

We have collected the majority of the runs and comparison between DEEP and JUWELS into Figure \ref{fig:all_runs} together with the modeled best-case performance for the MSA runs. We see that the GPUs significantly outperform the CPUs {\blue for Neko, similar to \citep{karp2022large},} while the strong scaling behavior when using GPUs is significantly worse. Scaling on the CPU clusters is nearly linear with a parallel efficiency between 90--110\% in almost all cases. The superlinear speedup we observe in for example the Pipe and TGV case on JUWELS is a well-known property of the spectral element method when strong scaling on multicore CPUs, this is discussed in for example~\cite{offermans2016strong}. For the GPUs, we achieve a parallel efficiency of 80\% for the first points while it decreases towards 50--60\% when we have $4000$ or fewer elements per GPU. We observed that in general, it was beneficial to put as many elements as possible on the GPU when in the extreme scaling domain, where the computing devices are close to their max capacity, due to their high performance. When in the scaling domain however, putting more elements on each CPU core gave the best performance, with each GPU computing around $60-120$ times the number of elements compared to a single CPU core.

Focusing first on the turbulent pipe case shown at the top of Figure \ref{fig:all_runs} we see how the performance is affected by distributing the computation between different computer architectures. As is proposed in the performance model, the performance is between the GPU and CPU performance and aligns well with the modeled line for the DEEP cluster. However, it becomes evident that the performance of this use case on JUWELS does not benefit from MSA, as the problem is small enough to be efficiently run on a single GPU node with four GPUs {\blue and 9000 elements per GPU. } The performance model for the best possible execution time follows a perfect linear scaling from one GPU. Therefore, it provides a very optimistic bound for JUWELS, {\blue significantly overpredicting the performance, because it does not take into account the impact of network communication when scaling beyond one GPU node. We also performed measurements using the local CPUs of the GPU nodes, but even in this case, the communication overhead surpassed the potential performance gain from using more CPUs on JUWELS. This can be partially explained by the vast imbalance between the GPU and CPU nodes on JUWELS, where the DDR memory of the host CPUs offers less than 10\% of the accumulated memory bandwidth of HBM memory on the GPUs. Partitioning the domain then leads to expensive memory transfers over PCIe.}

The primary case for MSA here would be when only one GPU is available, which cannot fit the entire problem. This is the case on the DEEP system. Using 1 GPU and 1 CPU node on DEEP results in more than $2\times$ speedup compared to 2 CPU nodes. Using both the GPU and CPU could thus potentially be beneficial for personal computers and desktops where the global memory of the GPU can not accommodate the entire problem. {\blue Of note is that the imbalance is lower on DEEP, as the number of GPUs per node is smaller. Using additional CPUs, both on the same node or another module yields here a proportionally larger performance improvement.}

For the TGV case, we see a similar performance curve to that of the turbulent pipe where the GPU and CPUs perform similarly. As for the MSA runs, we see that the performance for a few nodes is rather low as the CPUs need to carry a vast amount of memory, we are in other words limited by the $C_{max}$  of the GPUs, meaning that the CPUs must carry out the majority of the computational cost. For 16 computing devices, however, we find ourselves in the domain of our model where we can obtain additive performance in the best case as we scale up. For DEEP we get within 10--15\% of the best possible time for 8 and 16 computing devices, while we are within $10\%$ for 8 computing devices on JUWELS. For 32 on DEEP and 32--64 on JUWELS the communication time $T_c$ quickly impacts the performance we can achieve and the actual performance deviates more than 20\% from the modeled best-case, but the curve starts to align with the GPU-only scaling. We observe that for all cases up to 32 devices, the modeled performance predicts a worse performance than using the same number of GPU computing devices. For 64 devices the modeled performance of the MSA run would perform equally to the measured performance of 64 GPUs, assuming perfect scaling. At this point, however, the internal latency ($T_c$) of the computing units and communication overhead is significant, leading to a worse performance than modeled.

For the largest case, RBC, our results differ in some regards from the previous cases. As the Rayleigh-B\'{e}nard case has more than 2M elements, we cannot fit the problem on the DEEP Booster module {\blue where the GPUs only have 32~GB of HBM memory per GPU. We want to compare the number of computing devices between Cluster and Booster fairly, which prevented us from computing the problem with 48~GPUs, because the memory requirement is around 1~GB of memory per 1000 elements (meaning a total memory requirement of ~2000GB for the RBC case) for polynomial order 7. }As such we perform measurement only on the Cluster, comparing to the use of both the Booster and Cluster modules. The modeled best case then is based only on the best CPU performance and the computational cost dedicated to the CPUs. Here we can clearly see the opportunity of running a modular job to enable large problems to be efficiently executed. By using 48 GPUs in addition to 48 Cluster nodes, and using almost the whole DEEP system, we obtain a speedup of 2.7$\times$ compared to using almost the whole CPU module. {\blue However, one should note that the performance actually decreases compared to the Cluster-only runs when we execute the computation with 48 computing devices on DEEP. This is because of the lower memory capacity of the GPU nodes, which means that the number of elements per core is larger than when using only 48~CPUs. For 48 devices the number of elements increases from 1820 to 2400 as each V100 GPU can only accommodate a bit more than 30000 elements in the HBM memory. The cost $C$ per core then grows, and the runtime also increases, as predicted by our model. As such, one needs to consider that replacing one module with a high memory capacity by one with a higher performance and lower memory capacity does still decrease the cost per rank. Otherwise, the benefit of using a more powerful module does not improve the performance. This is no longer the case when using 48 GPUs: they then have a large enough capacity to also decrease the work per core for the CPUs.}

On JUWELS however, the performance increase is only prevalent for 64 computing devices, while using 64 GPUs + 64 CPU nodes gives a lower performance than only using 64 GPUs.  As such the primary benefit of inter-module jobs for CFD applications is in the domain when $T_c$ is comparably small and the Booster module does not have enough memory available to accommodate the problem. {\blue This corresponds to the extreme scale operation domain, which for Neko corresponds to when the number of elements (for polynomial order 7) is more than 20000, using half or more of the available HBM memory on the GPUs. It is only in this domain when additional computational resources are not as heavily affected by the different performance characteristics of the different modules and the performance is close to additive.}

\begin{figure}
\centering
 \begin{subfigure}[b]{0.49\textwidth}
\centering
    \includegraphics[width=\textwidth]{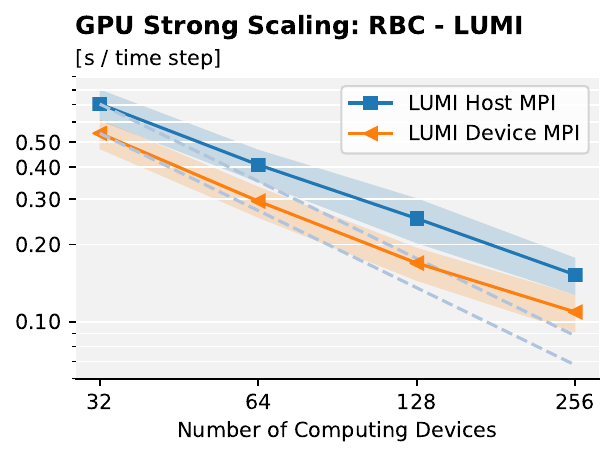}
    \label{fig:my_label}
 \end{subfigure} 
 \hfill
 \begin{subfigure}[b]{0.49\textwidth}
    \centering
    \includegraphics[width=\textwidth]{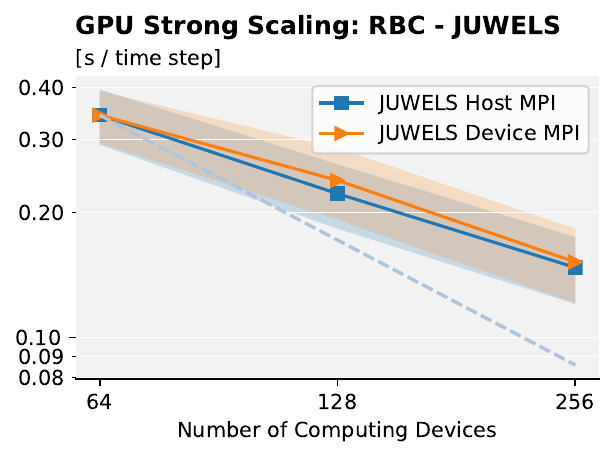}
    \label{fig:my_label}
 \end{subfigure}
 \caption{Performance comparison between the LUMI-G and JUWELS-Booster module where we compare utilizing the host for communication (host MPI) and utilizing device-aware MPI where the host is only used to schedule kernels on the device.}\label{fig:lumi-vs-juwels}
\end{figure}

 We also provide a comparison between the LUMI and JUWELS Booster modules for the RBC case in Figure \ref{fig:lumi-vs-juwels}. As we see in our measurements, CFD which can utilize GPUs is executed most efficiently on a large Booster-like system, we also provide this comparison between two current pre-exascale European supercomputers incorporating a modular design. As the best case in our measurements is to use the GPUs only to as large an extent as possible, we also include measurements with device-aware MPI enabled, where the MPI calls can be issued using pointers to memory on the device directly, further eliminating the host. One thing that is clear from the comparison between LUMI and JUWELS is that not using device-aware MPI on LUMI gives a significant performance penalty of 30--50\%, likely because the NIC is mounted on the GPU and as such using MPI on the host, leads to unnecessary data movement. For JUWELS, we observe a negligible difference between using device-aware MPI and host MPI, and it performs similarly to using host MPI on LUMI. Overall, one A100 performs better than one GCD of the MI250X when the number of nodes is small, but when the number of nodes is increased the improved network on LUMI makes up the difference. The difference between device-aware MPI and host MPI on JUWELS is smaller than 5\% and well within the standard deviation of a time step. This is in contrast with previous runs we executed using a mesh that was not load-balanced when device-MPI could perform as much as 6$\times$ better than using host MPI. These measurements indicate that if the problem is well partitioned, the usage of device vs host MPI is not as pronounced on JUWELS, but for ill-partitioned problems, the importance of device-aware MPI grows. On LUMI, not using device-aware MPI however, gives a significant performance penalty in all cases, performing 30--50\% worse than with device-aware MPI enabled. Compared to the CPU only runs on JUWELS, we observed that CPUs were much less affected by the partitioning of the elements between the different ranks. {\blue The reason for these large differences can be partially explained if we consider the node configuration on JUWELS Booster and LUMI-G and how the NICs are installed. For LUMI, they are mounted directly on the GPUs. This means that when device-aware MPI is not used, the data is first transferred to the host CPU and must pass through the GPU again before being communicated through the network. The same process is also applied when receiving messages. LUMI is as such not well suited for host-MPI. On JUWELS however, one would not expect the difference to be as pronounced as the PCIe switch is shared between the host and the GPU, and the data must not pass through the GPU an extra time when sending and receiving a message. However, still the difference compared to device-aware MPI is smaller than expected as one still executes two extra memory transfers to and from the CPU for each message. It is possible that the configuration of MPI we employ on JUWELS is not highly optimized for device-aware MPI. There is a significant number of options for the MPI runtime on JUWELS, e.g., by configuring UCX we would be able to achieve better use of device-aware MPI at this scale. In particular, during the runs on JUWELS the unreliable datagram (UD) setting with CUDA transport for UCX was used, intended for medium-sized simulations. It is possible that the low-memory DC (Dynamically Connected) option might be more performant at this scale. This option, however, had at the time of carrying out these experiments not been exhaustively tested on the JUWELS system.}

\subsection{Modeled Performance}
In our performance model, we are interested in modeling the best possible execution time given a set $S$ of computing devices. While we observed that it in some cases significantly overpredicts the performance for a mixed CPU-GPU run, it clearly illustrates how using only the strongest computing device to as large an extent as possible (assuming there are enough of them to accommodate the problem) is the way forward for large-scale homogeneous simulations. Although we have focused on CFD in our work, the same reasoning can be applied to any homogeneous workload where the main issue is to load balance parts of the problem between different ranks. As many applications fall in this category, our results support the trend of recent massively parallel systems to utilize primarily GPUs for the computation and dedicate a less powerful host only to schedule the computations. The latest candidates in this regard, LUMI, and Frontier, illustrate this trend clearly as the bandwidth and flop/s of the accelerators are more than $20 \times$ the performance of the host on a compute node. With upcoming architectures, we anticipate that the trend to remove the host from the computation and offload all tasks to the accelerator will continue. This is also the idea behind the Booster module where low-powered CPUs are equipped with powerful accelerators~(\citealt{Kreuzer:851724}). With this, we stress the point that for problems like CFD using a mix of CPU/GPU resources will likely not lead to any gains in the future, except for the case when the best-suited computing unit (in our case GPUs) cannot accommodate the entire problem. {\blue However, an opportunity is also to use applications such as this to backfill the computer resources when the system is idle. It is expected that incorporating more technologies such as malleable job-scheduling where jobs grow and shrink, applications that operate in the extreme-scale domain could use a simple performance model to indicate whether adding resources can be beneficial to decrease their time to solution. For Neko, in this case, our results indicate that the application operates in the extreme scale domain when more than half of the available memory is used on the GPUs. In a scenario where only some CPU resources are available directly to start the initialization of the problem, as more GPU resources become available the application accommodates more GPUs until the point at which the problem fits on only the Booster.}

\begin{figure*}
    \centering
    \begin{subfigure}[b]{0.495\textwidth}
\includegraphics[width=1\textwidth]{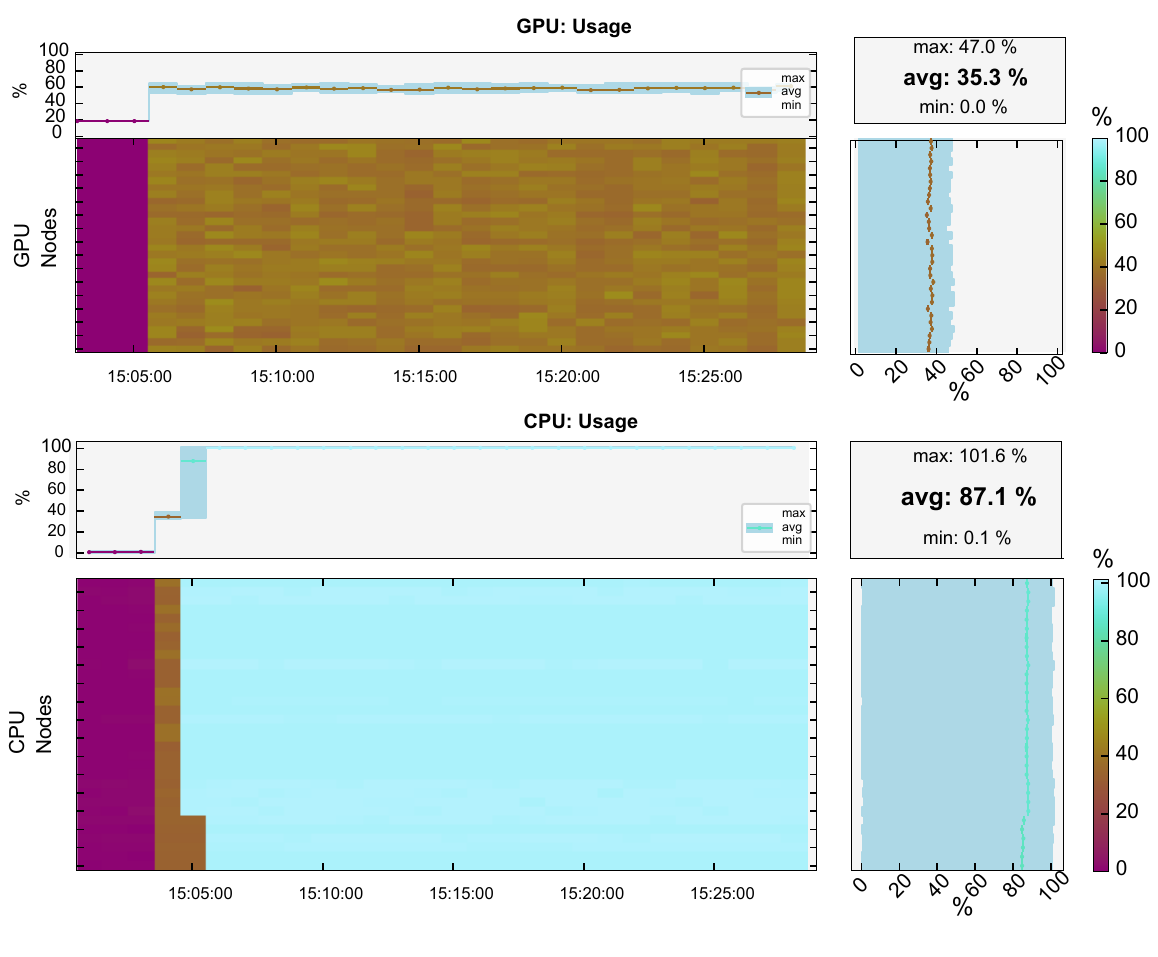}       
    \caption{}\label{subfig:noio}
    \end{subfigure}
    \hfill 
    \begin{subfigure}[b]{0.495\textwidth}
    \centering
\includegraphics[width=1\textwidth]{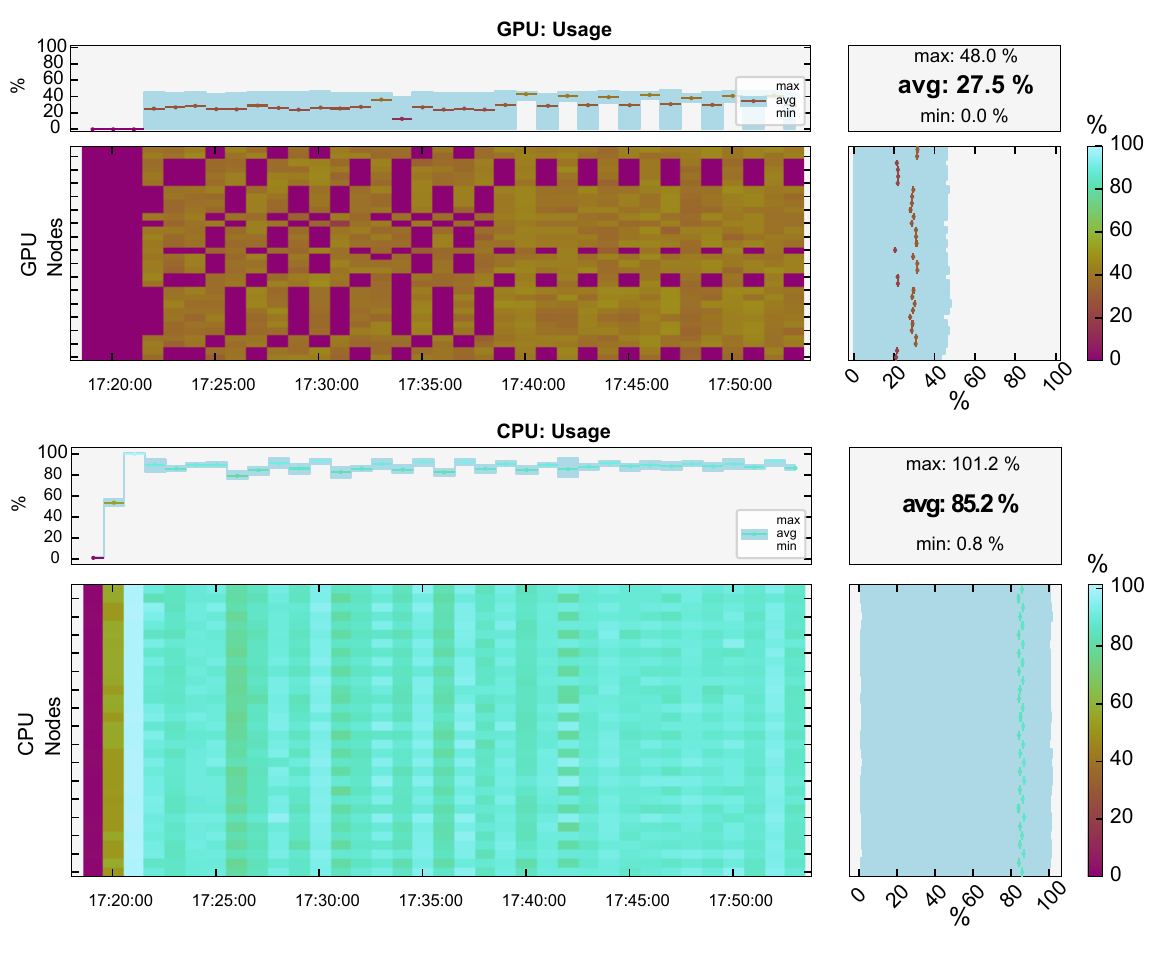}
    \caption{}\label{subfig:io}
    \centering
    \end{subfigure}
\caption{Performance traces with low-performance overhead from LLView for the GPU nodes (top) and CPU nodes (bottom) for an MSA run of the TGV case using 64 nodes split equally between GPU and CPU nodes (1:1 mix). The metric CPU and GPU usage are defined as the percent of time over the past sample period during which one or more kernels were executing on the GPU. In \textbf{\ref{subfig:noio}} a trace with no I/O. In \textbf{\ref{subfig:io}} a simulation with extensive I/O is presented.}
    \label{fig:LLview}
\end{figure*}
\subsection{I/O and mixing modules}
In the previous sections, we primarily considered the issue of balancing the load between different modules to the actual computation, however, for several applications I/O is the primary performance bottleneck. The impact of executing with a significant portion of I/O {\blue where output is written at each time step, vs one without any} I/O is shown from LLView in Figure \ref{fig:LLview}. From this, it is clear that not only must one then balance the computational load between the different computing devices, but also the writes to and from disk. The issue of balancing the load between devices can in the extreme case lead to a conflict between the computational load balance and the load on the file system. {\blue The I/O imbalance is due to the GPUs computing 100 times the number of elements compared to one CPU core, as such, on DEEP, this leads to the GPU nodes performing $100/24\approx 4$ times more I/O, greatly impacting the GPU usage.} This I/O imbalance leads to the GPUs spending a significant time idle compared to when not a lot of I/O is executed. The overall GPU utilization in this example is rather low though, as it is measured for the TGV case with 32 GPUs and 32 CPUs, and the problem size per computing device is comparably low. 

\section{Conclusions}
Our results support the notion that if the numerical method can both utilize CPUs and GPUs efficiently, executing large-scale CFD on a Booster-like system is beneficial when the problem fits on only this module. There is some room for improvement in the use of a mix of CPU and GPU nodes when the problem size is too large for the GPU module {\blue and when the HBM memory of the GPUs cannot fit the entire computational load, for our Neko setting this requirements was 1~GB of global memory per 1000 elements, but this may vary between cases and for other applications.  Overall, we observed that for this type of code where we utilize domain partitioning between the modules,} the communication overhead quickly becomes larger than the potential gain from using multiple computing devices. {\blue This is further amplified when a significant amount of I/O is carried out. As the GPUs have a higher performance and carry out a larger amount of work, they also write significantly more data to the parallel file system. While the performance of the GPUs is significantly higher, the bandwidth to disk is comparable to the CPU nodes, leading to a significant imbalance.} When the problem can fit on the GPUs only, it is best to utilize only the Booster, and even using the local host CPU gives a negligible or negative impact on the performance. For the GPU-only runs, we observe a difference between the JUWELS Booster and LUMI supercomputer when using device-aware MPI, primarily attributed to their respective network, {\blue and in particular to the NICs on LUMI being connected directly to the GPUs}. The performance of one Nvidia A100 on JUWELS is higher than LUMI's AMD MI250X GCD for a few nodes, but using device MPI improves the scaling on LUMI. We observe that the trend of moving to larger GPU-accelerated systems, where not only computation but also communication is offloaded to the most powerful computing units to increase locality, will benefit computational fluid dynamics applications able to efficiently offload the whole algorithm to the accelerator.

 
 
 


\begin{acks}
The authors gratefully acknowledge the computing time provided by the J\"{u}lich Supercomputing Centre (on JUWELS and DEEP). We acknowledge the National Academic Infrastructure for Supercomputing in Sweden (NAISS) and the Swedish National Infrastructure for Computing (SNIC) for awarding this project access to the LUMI supercomputer, owned by the EuroHPC-JU, hosted by CSC (Finland) and the LUMI consortium through a LUMI Sweden XLarge call. 
\end{acks}

\begin{dci}
The Author(s) declare(s) that there is no conflict of interest.
\end{dci}
\begin{sm}
    The Neko framework and the details for the test cases can be found on github.  The Neko package can be  downloaded here \url{https://github.com/ExtremeFLOW/neko} and the test cases on this link \url{https://github.com/ExtremeFLOW/MSA-tests}.
\end{sm}
\begin{funding}
The research in this paper has received funding from the European Union Horizon 2020  research and innovation programme under grant agreement No  955606 (DEEP-SEA). The EuroHPC Joint Undertaking (JU) receives support from the European Union Horizon 2020 research and innovation programme and Germany, France, Spain, Greece, Belgium, Sweden, Switzerland. Financial support was provided by the Swedish e-Science Research Centre Exascale Simulation Software Initiative (SESSI) and the Swedish Research Council project grant ''Efficient Algorithms for Exascale Computational Fluid Dynamics'' (grant reference 2019-04723).
\end{funding}
\bibliographystyle{sageh}

\end{document}